\documentclass{article} \usepackage{slashed,feynmf,epsfig,amsmath,amssymb,enumitem} \usepackage[papersize={8.5in,11in}]{geometry}
   \usepackage{bm}
\usepackage{graphicx}
\newcommand{\be}{\begin{equation}}
\newcommand{\ee}{\end{equation}}
\begin{document}
\title{Regular Rotating MOG Dark Compact Object}
\author{J. W. Moffat\\
Perimeter Institute for Theoretical Physics, Waterloo, Ontario N2L 2Y5, Canada}
\maketitle


\begin{abstract}
A regular rotating MOG compact object is derived that reduces to the Kerr black hole when the parameter $\alpha=0$. Physical consequences of the dark compact object, which is regular everywhere in spacetime for $\alpha > \alpha_{\rm crit}=0.674$ and is a rotating Kerr-MOG black hole for $\alpha < \alpha_{\rm crit}$ are investigated.
\end{abstract}

\maketitle


\section{Introduction}

The detection of gravitational waves by the LIGO/Virgo observatories has opened up a new possibility of investigating the nature of black holes~\cite{LIGO1,LIGO2,LIGO3,LIGO4}. The final merging of inspiralling binary black holes and neutron stars culminating in the ringdown phase and the formation of the remnant dark compact object can allow for distinguishing alternatives to black holes, such as quantum gravity and quantum effects induced dark compact objects and black holes~\cite{Cardoso}. Such exotic compact objects can not be completely dark and may not possess horizons and essential singularities. The black holes predicted to exist in general relativity are literally holes in spacetime, where classical physics breaks down at the essential singularity at the center of the black hole.  The existence of the one-way horizon membrane, causally disconnecting the black hole interior from its exterior, leads to conundrums such as the information loss problem~\cite{Hawking}. As more accurate data for the gravitational strain waveforms and the ringdown phase of merging binary dark compact objects are accumulated in the future, it will be possible to constrain alternatives to black holes.

In previous publications~\cite{Moffat1,Moffat2,Moffat3,Moffat4,MoffatManfredi} classical black hole solutions have been found in a modified gravity (MOG) theory~\cite{Moffat5}. The classical, static spherically symmetric Schwarzschild-MOG spacetime metric has two horizons, while the rotating Kerr-MOG black hole has two horizons and an ergosphere. In both these black hole solutions there is an essential singularity at the center of the black hole where the components of the Riemann tensor are singular. A static spherically symmetric solution of the MOG field equations for nonlinear field equations for the (spin 1 graviton) vector field $\phi_\mu$ and its field strength:
\be
B_{\mu\nu}=\partial_\mu\phi_\nu-\partial_\nu\phi_\mu
\ee
have been derived~\cite{Moffat2}. The thermodynamics properties of the MOG black holes and the regular MOG dark compact object have been investigated~\cite{Moffat4}. The quasinormal modes in the ringdown phase of the merging of two Schwarzschild-MOG black holes has been investigated~\cite{MoffatManfredi}. Other physical features of the MOG black holes and neutron stars have been investigated~\cite{Pradhan,Pradhan2,John,Hussein1,Hussein2,Armengol1,Armengol2,Armengol3,Sharif,Lee,Zakria,Sheoran,GreenMoffatToth,
Wei1,Wei2,Guo,Wondrak}. As an alternative gravity theory MOG has been shown to agree with data for galaxy rotation curves, galaxy dynamics and cosmology without dark matter~\cite{MoffatRahvar1,MoffatRahvar2,MoffatToth,BrownsteinMoffat,IsraelMoffat,MoffatCosmology,MoffatTothCosmology,Roshan,Moffat2020,MoffatII2020}.

The field equations for the matter-free MOG black hole metric spacetimes are given by
\be
R_{\mu\nu}=-8\pi GT_{\phi\mu\nu},
\ee
where we have set $c=1$ and assumed that the measure of gravitational coupling $G=G_N(1+\alpha)$ is constant, $\partial_\nu G = 0$, where $\alpha$ is a constant parameter, and the matter energy-momentum tensor $T_{M\mu\nu}=0$. We also need the matter-free field equations for the vanishing current density $J^\mu=0$:
\begin{equation}
\label{Bequation}
\nabla_\nu B^{\mu\nu}=\frac{1}{\sqrt{-g}}\partial_\nu(\sqrt{-g}B^{\mu\nu})=0,
\end{equation}
and
\begin{equation}
\label{Bcurleq}
\partial_\sigma B_{\mu\nu}+\partial_\mu B_{\nu\sigma}+\partial_\nu B_{\sigma\mu}=0,
\end{equation}
where $\nabla_\nu$ is the covariant derivative with respect to the metric tensor $g_{\mu\nu}$.

The gravitational energy-momentum tensor for the $\phi_\mu$ vector field is given by \footnote{We have assumed that the potential $V(\phi_\mu)$ in the definition in~\cite{Moffat1} of the energy-momentum tensor $T_{\phi\mu\nu}$ is zero.}
\begin{equation}
\label{Tphi}
T_{\phi\mu\nu}=-\frac{1}{4\pi}({B_\mu}^\alpha B_{\nu\alpha}-\frac{1}{4}g_{\mu\nu}B^{\alpha\beta}B_{\alpha\beta}).
\end{equation}
When the parameter $\alpha$ in the definition of the gravitational source charge $Q_g=\sqrt{\alpha G_N}M$ of the vector field $\phi_\mu$ vanishes, then the MOG field equations reduce to general relativity field equations. We neglect the mass of the $\phi_\mu$ field, for in the determination of galaxy rotation curves and galactic cluster dynamics $\mu=0.042\,(\rm kpc)^{-1}$, which corresponds to the vector field $\phi_\mu$ mass $m_\phi\sim 10^{-28}$ eV~\cite{MoffatRahvar1,MoffatRahvar2}. The smallness of the $\phi_\mu$ field mass in the present universe justifies our ignoring it when solving the field equations for compact objects such as neutron stars and black holes. It can be proved the mass $\mu$ of the vector field $\phi$ is zero for MOG black holes with horizons~\cite{MoffatII2020}.

The action for the above field equations is given by
\be
S_{\rm MOG}=\frac{1}{16\pi G}\int d^4x\sqrt{-g}[R-\frac{1}{4}B^{\mu\nu}B_{\mu\nu}],
\ee
where $R$ is the Ricci scalar.

A Kerr-MOG black hole or a regular MOG compact dark object can be formed from the collapse of a stellar object, and a supernassive Kerr-MOG black hole or regular supermassive MOG dark compact object can be formed by an as yet unknown mechanism. The gravitational repulsive force and pressure produced by the spin 1 gravitational vector field $\phi_\mu$ can prevent the formation of the dark compact object from becoming a MOG black hole. A regular solution for Einstein-Maxwell field equations with a nonlinear electrodynamics has been obtained~\cite{AyonBeatoGarcia}. After the formation of the dark compact object or black hole, an electrical charge $Q_e$ will be discharged extremely rapidly and the collapsed object will be electrically neutral~\cite{Carter,Gibbons,Wald}, so we ignore any contribution of electrical charge $Q_e$.

\section{Regular Static-MOG Compact Object}

The action for the matter-free MOG gravitational theory with nonlinear field equations for the gravitational spin 1 vector field $\phi_\mu$ is given by
\be
S_{\rm MOG}=\frac{1}{16\pi G}\int d^4x\sqrt{-g}[R-L(B)],
\ee
where $L(B)$ is the Lagrangian density describing the nonlinear contribution of $B_{\mu\nu}=\partial_\mu\phi_\nu-\partial_\nu\phi_\mu$ to the theory with $B=\frac{1}{4}B^{\mu\nu}B_{\mu\nu}$. The MOG field equations in the absence of matter are
\be
\label{Gequation}
G_{\mu\nu}=-8\pi GT_{\phi\mu\nu},
\ee
\be
\label{Bequation2}
\nabla_\nu(B^{\mu\nu}L_B)=0,
\ee
\be
\nabla_\nu(^*B^{\mu\nu})=0,
\ee
where $^*B^{\mu\nu}$ is the dual $B^{\mu\nu}$ tensor. The gravitational energy-momentum tensor is
\be
T_{\phi\mu\nu}=-\frac{1}{4\pi}[g_{\mu\nu}L(B)-B_{\mu\alpha}{B_\nu}^\alpha L_B],
\ee
where $L_B=dL_B/dB$.

As before, the gravitational constant $G=G_N(1+\alpha)$, or, for the units $G_N=1$, $G=1+\alpha$, and the gravitational source charge for the spin 1 gravitational vector field $\phi_\mu$ is $Q_g=\sqrt{\alpha}M$ where $M$ is the mass of the body. From the field equations, we obtain the gravi-electric field:
\be
E_g(r)=B_{01}(r)=-B_{10}(r).
\ee
This yields the energy-momentum tensor components:
\be
{T_\phi^0}_0={T_\phi^1}_1=-\frac{1}{4\pi}(L(B)+E_g^2L_B).
\ee

An alternative way to describe the nonlinear system is to consider the function $H$ obtained from the Legendre transformation~\cite{Plebanski}:
\be
H=2BL_B-L(B).
\ee
We define
\be
P_{\mu\nu}=L_BB_{\mu\nu},
\ee and
\be
P=\frac{1}{4}P_{\mu\nu}P^{\mu\nu}=(L_B)^2B,
\ee
where $H$ is a function of $P$.

The specific function $H$ for the regular spacetime metric solution is
\be
\label{H}
H(P)=P\frac{(1-3\sqrt{-2\alpha(1+\alpha)M^2P)}}{(1+\sqrt{-2\alpha(1+\alpha)M^2P)^3}}
-\frac{3}{2\alpha(1+\alpha)M^2b}\biggl(\frac{\sqrt{-2\alpha(1+\alpha)M^2P}}{1+\sqrt{-2\alpha(1+\alpha)M^2P}}\biggr)^{3/2},
\ee
where $b=\sqrt{\alpha}M/2$ and $P=-\alpha(1+\alpha)M^2/2r^4$. The corresponding Lagrangian $L$ is given by
\be
L(P)=P\frac{(1-8\sqrt{-2\alpha(1+\alpha)M^2P}-6\alpha(1+\alpha)M^2P)}{(1+\sqrt{-2\alpha(1+\alpha)M^2P})^4}
-\frac{3(-2\alpha(1+\alpha)M^2P)^{5/4}(3-2\sqrt{-2\alpha(1+\alpha)M^2P})}{4\alpha(1+\alpha)M^2b(1+\sqrt{-2\alpha(1+\alpha)M^2P})^{7/2}}.
\ee

\section{Regular Static Spherically Symmetric Compact Object}

For the static spherically symmetric line element, we have
\be
\label{metric}
ds^2=f_S(r)dt^2-{f_S(r)}^{-1}dr^2-r^2d\Omega^2,
\ee
where $d\Omega^2=d\theta^2+\sin^2\theta d\phi^2$. The exact $f_S(r)$ for the MOG regular static, spherically symmetric dark compact object satisfying the weak energy condition is given by~\cite{AyonBeatoGarcia,Moffat2}:
\be
\label{fSmetric}
f_S(r)=1-\frac{2(1+\alpha)Mr^2}{(r^2+\alpha(1+\alpha)M^2)^{3/2}}+\frac{\alpha(1+\alpha)M^2r^2}{(r^2+\alpha(1+\alpha)M^2)^2}.
\ee
The gravi-electric field is given by
\be
E_g(r)=\sqrt{\alpha}Mr^4\biggl(\frac{r^2-5\alpha(1+\alpha)M^2}{(r^2+\alpha(1+\alpha)M^2)^4}
+\frac{15}{2}\frac{(1+\alpha)M}{(r^2+\alpha(1+\alpha)M^2)^{7/2}}\biggr).
\ee
We obtain for large $r$ the asymptotic behavior:
\be
f_S(r)\approx 1-\frac{2(1+\alpha)M}{r}+\frac{\alpha(1+\alpha)M^2}{r^2}.
\ee

Near the center $r\approx 0$ the static, spherically symmetric dark compact object described by the metric (\ref{fSmetric}) has the behavior:
\be
f_S(r)\equiv g_{00}(r)\approx 1-\frac{1}{3}Cr^2,
\ee
where $C$ is given by
\be
C=\frac{3}{M^2}\biggl[\frac{2(1+\alpha)^{1/2} - \alpha^{1/2}}{\alpha^{3/2}(1+\alpha)}\biggr],
\ee
and the spacetime metric is regular $f_S(0)=1$. The curvature invariants $R={R^\mu}_\mu$ and $R^{\mu\nu\lambda\sigma}R_{\mu\nu\lambda\sigma}$ are regular at r=0.

let us rewrite Eqs.(\ref{Gequation}) and (\ref{Bequation}) in the form~\cite{AyonBeatoGarcia}:
\be
\label{Gequat}
G_\mu^\nu=2(H_PP_{\mu\lambda}P^{\mu\lambda}-\delta_\mu^\nu(2PH-H),
\ee
\be
\nabla_\nu P^{\mu\nu}=0,
\ee
where
\be
P_{\mu\nu}=-2\delta_{[\mu}^0\delta_{\nu]}^r\frac{\sqrt{\alpha(1+\alpha)}M}{r^2},\quad P=-\frac{\alpha(1+\alpha)M^2}{2r^4}.
\ee
The $G_0^0$ in equation (\ref{Gequat}) yields:
\be
\frac{rdA/dr-A}{r^4}=2H(P).
\ee
By substituting $H$ from (\ref{H}), we obtain after an integration;
\be
A(r)=2(1+\alpha)Mr-\frac{2(1+\alpha)Mr^4}{(r^2+\alpha(1+\alpha)M^2)^{3/2}}+\frac{\alpha(1+\alpha)M^2r^4}{(r^2+\alpha(1+\alpha)M^2)^2}.
\ee
Substituting $A(r)$ into:
\be
f_S(r)=1-\frac{2(1+\alpha)M}{r}+\frac{A(r)}{r^2},
\ee
we get the basic solution (\ref{fSmetric}).

To remove the singular coordinate behavior of the metric, we rewrite the metric in advanced Eddington-Finkelstein coordinates. We perform the following transformation for an incoming photon:
\be
v=t-r^*,
\ee
and for an outgoing photon:
\be
\label{outgoingphoton}
u=t+r^*,
\ee
where
\be
r^*=\int\frac{dr}{f_S(r)}.
\ee
The metric in the advanced Eddington-Finkelstein coordinates is of the form:
\be
\label{EFmetric}
ds^2=f_S(r)du^2+2dudt-r^2d\Omega^2.
\ee

The gravitational redshift $z$ is determined for the static spherically symmetric system by
\be
z(r)=\frac{1}{\sqrt{f_S(R)}}-1,
\ee
where $R$ is the radius of the compact object and $r$ is the asymptotic distance to an observer. When $Q_g\leq Q_{g\rm crit}$ and $\alpha\leq
\alpha_{\rm crit}=0.674$, the redshift is infinite at the horizon $r_+$ where
\be
\label{horizons}
r_\pm=M[1+\alpha\pm(1+\alpha)^{1/2}],
\ee
while for $\alpha > \alpha_{\rm crit}=0.674$, the redshift $z$ is finite. We expect observationally that the regular dark compact object is sufficiently dark to be compatible with binary x-ray observations, so $\alpha\sim \alpha_{\rm crit}$. In Fig. 1, we display the redshift as a function of $\alpha$. The redshift approaches infinity as $\alpha\rightarrow \alpha_{\rm crit}=0.674$. In Figure 2, the ratio of $r(z_{\rm max})/G_N(1+\alpha)M$ versus $\alpha$ is displayed.

\begin{figure}
\centering\includegraphics[scale=0.8]{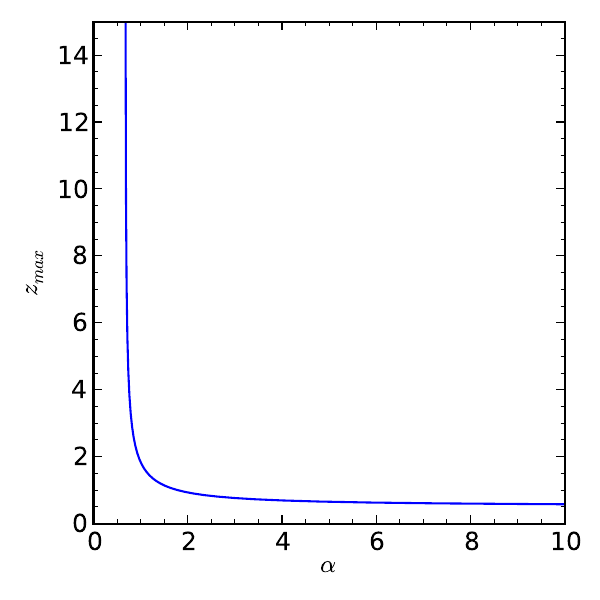}\\
\caption{Redshift versus $\alpha$.}{\label{fig.z_maxVSalpha}}
\end{figure}

\begin{figure}
\centering\includegraphics[scale=0.8]{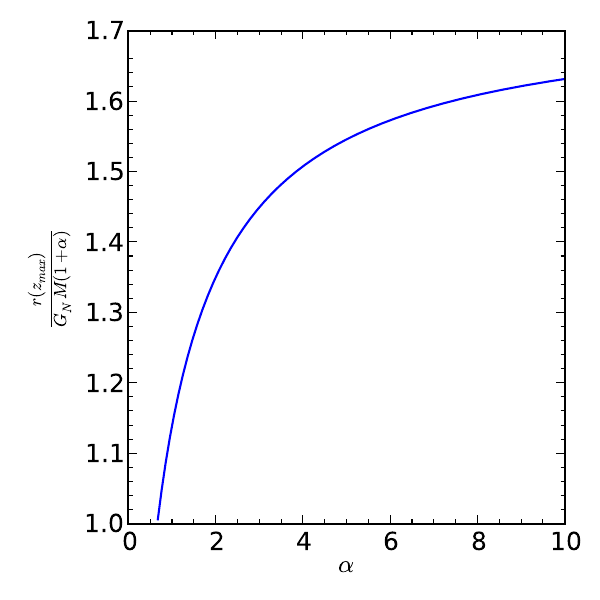}\\
\caption{The ratio $r(z_{\rm max})/G_N(1+\alpha)M$ versus $\alpha$.}{\label{fig.r_zmaxVSalpha}}
\end{figure}

The repulsive {\it gravitational} force and pressure $p$, produced by the spin 1 (graviton) vector field $\phi_\mu$, is responsible for the prevention of the collapse of a star to a black hole with horizons and an essential singularity at $r=0$.

\section{Regular MOG Rotating Compact Object}

We will use the Newman-Janis method to convert the spherically symmetric, static compact object into a rotating one~\cite{JanisNewman,Toshmatov}. The Janis-Newman algorithm in its original formulation generates rotating metric solutions from static metric solutions. It was found to provide an alternative derivation of the Kerr metric solution~\cite{Kerr} of Einstein's field equations. The Janis-=Newman algorithm generates axisymmetric metrics from a spherically symmetric metric through the choice of the complexification of radial and null time coordinates. This is followed by a complex coordinate transformation. Finally a change of real coordinates is performed to write the result in Boyer-Lindquist coordinates. Newman and Janis used the Newman-Penrose tetrad formalism, which required inverting the metric to find a null tetrad basis used to perform the applied transformation, and lastly to again invert the metric. The transformations can be well defined by showing that the algorithm is consistent with a class of Kerr-Schild metrics.

We begin by considering only the outgoing photon case (\ref{outgoingphoton}) and make the following complex coordinate transformations:
\be
{\tilde r}=r+ia\cos\theta,\quad {\tilde u}=u-ia\cos\theta,\quad {\tilde\theta}=\theta,\quad {\tilde\phi}=\phi,
\ee
where $a=J/M$ and $J$ denotes the angular momentum. From these transformations, we obtain the metric function:
\be
\label{Rotmetric}
f_R(r)=1-\frac{2(1+\alpha)Mr\rho}{(\rho^2+\alpha(1+\alpha)M^2)^{3/2}}+\frac{\alpha(1+\alpha)M^2\rho^2}
{(\rho^2+\alpha(1+\alpha)M^2)^2},
\ee
where $\rho^2=r^2+a^2\cos^2\theta$. The metric line element now becomes in the Boyer-Lindquist coordinates:
\be
ds^2=f_R(r)dt^2-\frac{\rho^2dr^2}{\rho f_R(r)+a^2\sin^2\theta}+2a\sin^2\theta[1-f_R(r)]d\phi dt-\rho^2d\theta^2
-\sin^2\theta[\rho^2-a^2[f_R(r)-2]\sin^2\theta]d\phi^2.
\ee
The rotating MOG solution is fully determined by the mass $M$, the spin parameter $a$ and the parameter $\alpha$.

The Kerr-MOG metric line element becomes the Kerr metric line element when $\alpha=0$~\cite{Kerr}
:
\be
\label{Kerrmetric}
ds^2=\biggl(1-\frac{2Mr}{\rho^2}\biggr)dt^2-\frac{\rho^2}{\Delta}dr^2+2(2Mr){\rho^2}a\sin^2\theta d\phi dt
-\rho^2s\theta^2-\biggl(r^2+a^2+\frac{2Ma^2r\sin^2\theta}{\rho^2}\biggr)\sin^2\theta d\phi^2,
\ee
where $\Delta=r^2+a^2-2Mr$.

There is a critical value of the gravitational charge $Q_g=Q_{g\rm crit}$ when the regular rotating MOG dark compact object becomes a rotating MOG black hole with two horizons and an ergosphere. The critical value $Q_{g\rm crit}$ is obtained when the two conditions are satisfies:
\be
g_{00}(r,a,\theta,\alpha)=0,\quad \partial_r g_{00}(r,a,\theta,\alpha)=0.
\ee
Solving these equations for $r$ and $\alpha$, we can obtain a solution as a function of $a$, $\theta$ and $\alpha$:
\be
g_{rr}(r,a,\theta,\alpha)=0,\quad \partial_r g_{rr}(r,a,\theta,\alpha)=0.
\ee
The critical values of the gravitational charge $Q_{g\rm crit}$ and $\alpha_{\rm crit}$ are determined by the radius of the static limit surface $r$ corresponding to the rotation parameter $a$ for different values of $\theta$. By solving for the roots of $f_R(r,\alpha)=g_{00}(r,\alpha)=0$, we obtain the critical value $\alpha=\alpha_{\rm crit}\leq 0.674$, which separates the MOG black hole from the regular MOG dark compact object. For $\alpha\leq \alpha_{\rm crit}=0.674$ the static black hole with $a=0$ has the two horizons (\ref{horizons}).

For $\alpha_{\rm crit} > 0.674$ the static, dark compact object is regular throughout spacetime with no horizons and no singularity at $r=0$. For the rotating MOG compact object the critical value $\alpha_{\rm crit}$ for a range of $a$ and $\theta$ physically constrains the compact object to two possible physical systems. For $\alpha < \alpha_{\rm crit}$ the rotating compact object has two horizons and an ergosphere, while for $\alpha > \alpha_{\rm crit}$ there are no horizons and there can and cannot be an ergosphere depending on the value of the coordinate $\theta$. The rotating MOG dark compact object without horizons, and no singularity at $r=0$ is regular everywhere in spacetime.

\section{Conclusions}

We have derived a generalized Kerr rotating dark compact object determined by the angular momentum (spin), mass and the parameter $\alpha$ of the object. When $\alpha > \alpha_{\rm crit}=0.674$, the static compact object and MOG spinning compact object can, depending on the spin parameter $a$ and coordinate $\theta$, be regular without horizons and with and without an ergosphere. The repulsive gravitational force and pressure, preventing the compact object from becoming a rotating Kerr-MOG black hole with $\alpha < \alpha_{\rm crit}$, are furnished by the gravitational vector $\phi_\mu$ and its corresponding field strength $B_{\mu\nu}$. The LIGO/Virgo observatory data can provide constraints on the alternative dark compact object. In particular, future accurate data for the waveforms of merging binary dark compact objects can distinguish between a general relativity black hole, a MOG black hole and the static and rotating MOG horizonless and singularity-free dark compact object.

In the event that the MOG regular dark compact object can be a physical alternative to a black hole, then this would have the important consequence of removing black hole conundrums such as the information loss paradox~\cite{Hawking}. It would demonstrate that it is possible to describe the regular spacetime properties of dark compact objects by classical gravity theory. In classical general relativity the matter in the collapse of a star to a black hole ends up residing at the essential singularity at the center of the black hole. In the case of a MOG regular dark compact object, the matter would reside inside the compact object as a regular interior solution of the field equations, including a matter energy-momentum tensor $T_{M\mu\nu}$. The stability of the massive compact object would be determined by solving generalized MOG Oppenheimer-Volkoff equations with an equation of state for the matter fluid compatible with the stabilizing, MOG gravitational repulsive force and pressure. Investigations have been performed to constrain material binary compact objects using LIGO/Virgo gravitational wave data and simulated wave forms~\cite{Giudice,Khan,McDaniel}.

Is there a law of nature that could inhibit the collapse of a star to a MOG black hole by compelling $\alpha > \alpha_{\rm crit} = 0.674$? The answer to this question requires a solution of the physical determination in MOG of the parameter $\alpha$, or, in turn, a MOG determination of the physical behavior of $G=G_N(1+\alpha)$ and the gravitational charge $Q_g=\sqrt{\alpha G_N}$ for weak as well as strong gravitational fields.

\section*{Acknowledgments}

I thank Martin Green and Viktor Toth for helpful discussions, and I thank Martin Green for providing Figures 1 and 2. This research was supported in part by Perimeter Institute for Theoretical Physics. Research at Perimeter Institute is supported by the Government of Canada through the Department of Innovation, Science and Economic Development Canada and by the Province of Ontario through the Ministry of Research, Innovation and Science.

\end{document}